\documentclass[letterpaper]{article} 
\usepackage{aaai2026}  
\usepackage{times}  
\usepackage{helvet}  
\usepackage{courier}  
\usepackage[hyphens]{url}  
\usepackage{graphicx} 
\usepackage{natbib}  
\usepackage{caption} 
\usepackage{algorithm}
\usepackage{algorithmic}

\usepackage{newfloat}
\usepackage{listings}
\usepackage{amssymb}
\usepackage{amsmath}
\usepackage{booktabs}
\usepackage{multirow}
\DeclareCaptionStyle{ruled}{labelfont=normalfont,labelsep=colon,strut=off} 
\floatstyle{ruled}
\newfloat{listing}{tb}{lst}{}
\floatname{listing}{Listing}
\title{Speech Recognition Model Improves Text-to-Speech Synthesis using \\ Fine-Grained Reward}
\author{
    Guansu Wang\textsuperscript{\rm 1}, Peijie Sun\textsuperscript{\rm 2}\thanks{Corresponding author.}
}
\affiliations{
    \textsuperscript{\rm 1}University of Melbourne, Melbourne, Australia\\

 \textsuperscript{\rm 2}Nanjing University of Posts and Telecommunications\\
}

\usepackage{bibentry}

\begin{document}

\maketitle

\begin{abstract}
Recent advancements in Text-to-Speech (TTS) technology have been remarkable, enabling current models to clone arbitrary unseen speakers and synthesize high-quality, natural-sounding speech. However, corresponding evaluation techniques appear to be lagging: Existing Mean Opinion Score (MOS) estimation models typically perform regression-based scoring on entire speech segments--while a failed synthesized speech usually contains problematic elements in only a few isolated words rather than throughout the entire utterance. In this context, we presents an intriguing finding: encoder-decoder ASR models, such as Whisper, leverage their extensive pre-training to precisely capture word-level mismatches between speech and text within their cross-attention mechanisms, thereby providing a fine-grained reward signal. Building upon this insight, we propose a novel TTS optimization method, which we term \textbf{W}ord-level TTS \textbf{A}lignment by \textbf{A}SR-driven \textbf{A}ttentive \textbf{R}eward (W3AR). Instead of relying on any explicit reward annotations, W3AR leverages the attention information within a pre-trained ASR model, enabling finer-grained alignment and optimization of the sequences predicted by the TTS model. Experimental results demonstrate that W3AR not only effectively improves the TTS generation quality of existing models but also further enhances zero-shot robustness based on both in-domain and out-of-domain prompt speakers. Additionally, our findings and proposed methodology offer a new insight for generative tasks: understanding models can potentially serve as evaluators, providing highly fine-grained and valuable feedback for generation.
\end{abstract}


\section{Introduction}
The current wave of Artificial Intelligence-Generated Content (AIGC) has profoundly impacted various domains, with Text-to-Speech (TTS) technology standing out as a pivotal component~\cite{cao2025survey}. However, generating high-fidelity, high-sampling-rate signals such as human speech presents unique and substantial challenges~\cite{qian2014training}. In recent years, remarkable progress has been achieved, largely attributable to advancements in speech tokenization techniques and the availability of speech-text paired datasets~\cite{wang2023neural}. Existing TTS models now exhibit impressive zero-shot capabilities, notably the ability to perform voice cloning, synthesizing high-quality speech content in the voice of any unseen speaker~\cite{chen2024vall,wang2024maskgct}. \par
The recent breakthroughs in TTS can largely be attributed to two primary technical paradigms: diffusion-based models~\cite{popov2021grad} and autoregressive models~\cite{wang2023neural}. Despite of both remarkable performance, this paper primarily focuses on the latter due to its inherent advantages. Autoregressive models generate speech sequences by sequentially predicting tokens, allowing for dynamic determination of output length and generally supporting faster, streaming inference~\cite{du2024cosyvoice2}. Furthermore, the discrete tokens generated by these models can be subsequently refined with richer acoustic details, often through techniques like flow matching~\cite{du2024cosyvoice}. However, despite their strengths, autoregressive TTS systems are prone to certain generation artifacts. Specifically, synthesized speech segments can sometimes suffer from misspoken words, word repetitions, or acoustic distortions~\cite{xin2024rall}. These issues often stem from the cumulative error propagation inherent in sequential generation, where errors in early token predictions can cascade and amplify through the sequence, leading to noticeable imperfections in the final utterance~\cite{neekhara2024improving}. \par
Some existing efforts have explored RL as a potential avenue for posterior optimization in TTS, demonstrating its promise in refining generated speech~\cite{zhang2024speechalign}. However, the most primary challenge in this paradigm lies in obtaining a suitable reward signal. Standard TTS evaluation metrics, such as Mean Opinion Score (MOS) estimation, provide a holistic judgment for an entire utterance, lacking the fine-grained granularity required for effective RL. For instance, if only a short segment (e.g., 1 or 2 words) of an otherwise high-quality speech sequence contains an artifact, penalizing the entire sequence's probability will inevitably diminish optimization efficiency~\cite{yao2025fine}. Consequently, our research addresses the critical question of how to precisely locate these problematic segments to provide more effective and targeted optimization. \par
Considering that ASR models are inherently designed to learn the ``alignment" between two disparate modalities, i.e., speech and text, of varying lengths, we posit a direct hypothesis: can this alignment serve as a word-level evaluation metric to reflect the quality of synthesized spoken words? Prior research has already demonstrated the rich information embedded within the attention matrices of large pre-trained models~\cite{ben2024attend,hu2024self}. Building on this, we propose a novel metric based on the cross-attention mechanism within ASR models to assess the word-level alignment between generated speech and the given text from an ASR perspective. It is crucial to note that such a metric offers more comprehensive information than a simple Word Error Rate (WER). Even for ambiguously generated speech segments, an ASR model can often provide a robust interpretation by leveraging contextual information, which a simple WER might misclassify as an error. \par
To this end, we introduce W3AR, a novel framework for Word-level, Whisper-guided Audio Refinement using Adversarial Rewards. Our approach operationalizes the ASR alignment hypothesis by defining two fine-grained quality metrics. The first, Attention Purity, evaluates the articulatory clarity of individual words by assessing whether the ASR model's attention is sharply focused on a compact audio segment or diffusely scattered, indicating ambiguity. The second, Alignment Monotonicity, assesses the prosodic fluency of the utterance by ensuring the ASR's attention focus progresses smoothly forward in time, penalizing unnatural stalls or regressions that correspond to stutters or awkward pauses. These metrics are combined into a word-level reward signal that guides the optimization of the TTS model within a stable, group-relative policy optimization framework, directly targeting and correcting specific generation artifacts. \par

Our extensive experiments demonstrate the effectiveness and generality of W3AR. When applied to the state-of-the-art CoSyVoice~\cite{du2024cosyvoice2} model, our method yields significant reductions in Word Error Rate and notable increases in both objective speaker similarity and subjective Mean Opinion Scores for naturalness. Crucially, we show that these improvements hold not only for speakers within the training distribution but also for challenging out-of-domain speakers, confirming the robustness of our approach. Furthermore, we validate the model-agnostic nature of W3AR by successfully applying it to other diverse TTS architectures, including VoiceCraft and MaskGCT~\cite{wang2024maskgct}, achieving consistent performance gains. Ablation studies confirm that both the purity and monotonicity components are essential for achieving optimal results. \par

In summary, our main contributions are threefold: (1) We propose a novel, fine-grained reward function for TTS based on analyzing the purity and monotonicity of cross-attention maps from a pre-trained ASR model. (2) We design an effective and stable policy optimization framework, W3AR, that leverages this reward to directly correct word-level defects in synthesized speech. (3) We provide a comprehensive empirical validation of our method's effectiveness and generality, demonstrating significant improvements across multiple state-of-the-art TTS models and on both in-domain and out-of-domain datasets.
\section{Related Work}
\subsection{Zero-shot TTS}
Recent advancements in large-scale generative modeling have catalyzed a significant paradigm shift within the field of text-to-speech (TTS). A prevailing trend involves reformulating speech synthesis as a next-token prediction problem~\cite{wang2023neural}, mirroring the successes observed with LLMs in the text domain~\cite{peng2024voicecraft}. This approach fundamentally relies on the use of a neural audio codec to discretize continuous speech waveforms into a sequence of discrete tokens~\cite{zeghidour2021soundstream,wu2023audiodec}. A powerful decoder-only language model is then conditioned on phonetic or textual inputs to autoregressively predict this stream of acoustic tokens~\cite{du2025vall}. Pioneering work in this area, notably VALL-E~\cite{wang2023neural}, first demonstrated the profound potential of this methodology. This generative framework has proven to be exceptionally scalable, demonstrating that increasing model and dataset size yields substantial gains in output quality~\cite{anastassiou2024seed}, robustness~\cite{wang2025spark}, and the ability to capture nuanced prosodic details for any target voice without explicit fine-tuning.
\subsection{Alignment in TTS}
The alignment of generative models with human preferences has become a cornerstone of contemporary AI research, with Reinforcement Learning from Human Feedback (RLHF) emerging as an instrumental technology~\cite{zheng2023secrets}. This paradigm was first prominently established in Natural Language Processing (NLP)~\cite{kaufmann2024survey}, where preference optimization conventionally involves maximizing a reward signal produced by a separately trained reward model~\cite{lee2023rlaif,wu2023fine}. \par
In contrast to its maturity in NLP, the application of RLHF to TTS is a more nascent yet rapidly advancing frontier. Initial explorations like SpeechAlign~\cite{zhang2024speechalign} first adapted preference alignment to TTS using paired comparison data. Subsequent research has broadened this scope; for instance, UNO~\cite{chen2024enhancing} and RIO~\cite{hu2024robust} were developed to handle more complex, unpaired preference datasets by accounting for annotation uncertainties and employing Bayesian-inspired data selection strategies, respectively. While empirical studies and applications like Seed-TTS~\cite{anastassiou2024seed} have validated the efficacy of RLHF for enhancing the quality of LM-based TTS during post-training, a significant limitation persists. The predominant focus of current methodologies remains on coarse, utterance-level preference optimization~\cite{tian2025preference}, largely overlooking the substantial potential that lies in achieving more fine-grained acoustic alignment.
\section{Method}
\begin{figure*}[h!]
\centering
\includegraphics[width=1.295\columnwidth]{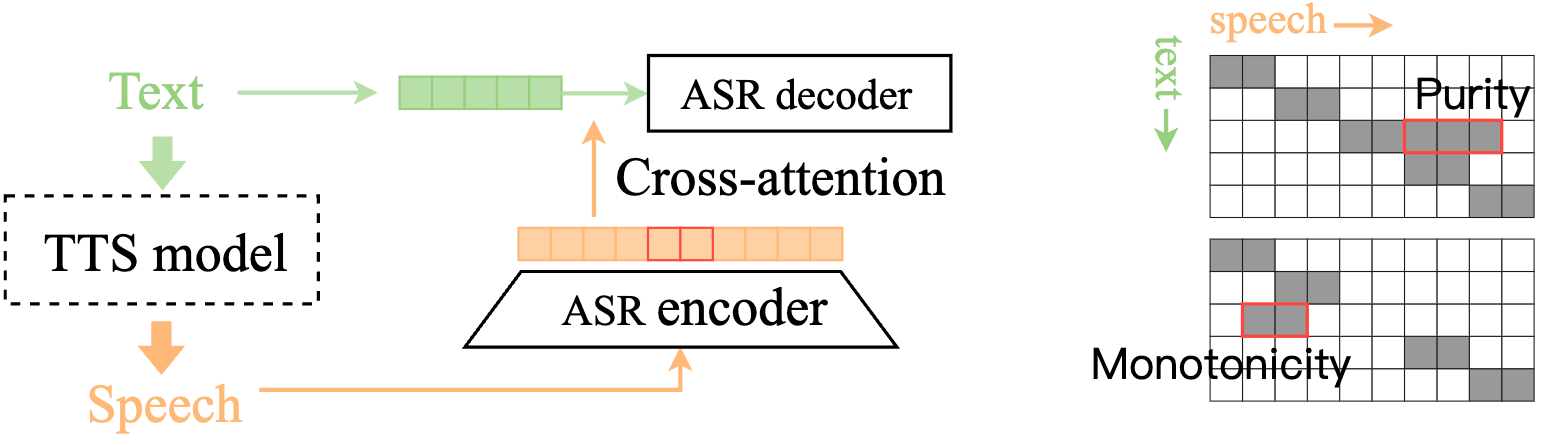} 
\caption{An overview of our proposed W3AR framework. The left panel illustrates the core process: speech synthesized by a TTS model is fed into the ASR encoder, while the ground-truth text guides the ASR decoder. The cross-attention mechanism acts as a bridge, where the ASR decoder uses each text token as a query to probe the acoustic representations from the encoder. The right panel visually defines the failure modes captured by our two metrics: (Top) Low Purity, where attention for a single token is diffuse and scattered across many audio frames, indicating unclear articulation. (Bottom) Poor Monotonicity, where the attention peak stalls or regresses, indicating unnatural prosody or rhythm.}
\label{fig1}
\end{figure*}
\subsection{Foundation: Zero-shot Autoregressive TTS}
We build upon a modern zero-shot TTS framework that typically leverages autoregressive language model to generate speech from discrete acoustic representations. This approach enables voice cloning from a short, unseen audio prompt without requiring speaker-specific fine-tuning. \par
\noindent\textbf{Discrete Speech Representation.} We utilize a pre-trained neural audio codec, denoted by $\mathcal{C}$, to transform a continuous audio waveform $\mathbf{x} \in \mathbb{R}^{T_u}$ into a sequence of discrete acoustic tokens $\mathbf{a} = (\mathbf{a}_1, \dots, \mathbf{a}_{T_a})$. The codec's quantizer employs $K$ hierarchical codebooks, so each token $\mathbf{a}_t$ is a stack of $K$ integer indices: $\mathbf{a}_t = (a_t^1, \dots, a_t^K)$, where $a_t^k$ is an index from the $k$-th codebook. The codec's parameters are frozen throughout our experiments.

\noindent\textbf{Model Training.}
The model $\mathcal{M}_\theta$ is trained to predict the acoustic token sequence $\mathbf{a}$ conditioned on the corresponding input text sequence $\mathbf{y} = (y_1, \dots, y_{T_y})$. For a given training pair $(\mathbf{y}, \mathbf{a})$ from a large, multi-speaker dataset $\mathcal{D}$, the model is optimized to maximize the likelihood of the acoustic sequence via a standard cross-entropy loss:
\begin{equation}
    \mathcal{L}_{\text{train}} = -\mathbb{E}_{(\mathbf{y}, \mathbf{a}) \sim \mathcal{D}} \left[ \sum_{t=1}^{T_a} \log p(\mathbf{a}_t \mid \mathbf{a}_{<t}, \mathbf{y}; \theta) \right]
\end{equation}
The prediction of the token stack $\mathbf{a}_t$ is factorized in a coarse-to-fine manner, predicting the primary token $a_t^1$ first, followed by the remaining detail tokens:
\begin{equation}
    p(\mathbf{a}_t \mid \cdot) = p(a_t^1 \mid \cdot) \prod_{k=2}^{K} p(a_t^k \mid a_t^1, \cdot)
\end{equation}
This process can be implemented using different heads~\cite{peng2024voicecraft} or predicted in a non-autoregressive manner~\cite{chen2024vall}. Moreover, flow-matching enables high-quality mel-spectrogram prediction from a single codebook layer~\cite{du2024cosyvoice}.

\noindent\textbf{Zero-Shot Voice Cloning via In-Context Learning.}
To enable zero-shot capabilities, we frame the task as speech continuation. During training, we define an acoustic prompt $\mathbf{a}_{\text{prompt}} = (\mathbf{a}_1, \dots, \mathbf{a}_{T_p})$, which is a short prefix (e.g., 3 seconds) of the full acoustic sequence $\mathbf{a}$. The model is then trained to predict the remainder of the sequence, $\mathbf{a}_{\text{target}} = (\mathbf{a}_{T_p+1}, \dots, \mathbf{a}_{T_a})$, conditioned on both the text $\mathbf{y}$ and the prompt $\mathbf{a}_{\text{prompt}}$. The objective implicitly teaches the model to infer the vocal characteristics from the prompt and maintain them throughout the subsequent generation. During inference, we provide the model with a prompt $\mathbf{a}_{\text{prompt}}$ from any unseen target speaker and a new text $\mathbf{y}$. The model autoregressively generates the acoustic sequence $\mathbf{a}_{\text{gen}}$ by sampling from the learned distribution $p(\cdot \mid \mathbf{a}_{\text{prompt}}, \mathbf{y}; \theta)$. The self-attention mechanism of the Transformer enables the model to attend to the acoustic properties embedded in the prompt, thus achieving zero-shot voice cloning even without speaker-specific fine-tuning.

\subsection{ASR-driven Reward Modeling}
To quantitatively assess the quality of the synthesized speech, we employ a pre-trained, frozen encoder-decoder ASR model, $\mathcal{M}_{\text{ASR}}$, architecturally similar to Whisper~\cite{radford2023robust}. This model serves as an objective evaluator, providing a fine-grained reward signal derived from its internal cross-attention mechanism.

The ASR model’s audio encoder, $\mathcal{E}_{\text{ASR}}$, maps a log-Mel spectrogram of the input audio into a sequence of hidden representations $\mathbf{H} \in \mathbb{R}^{T_h \times D}$. During the transcription process, the ASR’s text decoder, $\mathcal{D}_{\text{ASR}}$, predicts a sequence of text tokens. For each predicted text token $y_t$, a cross-attention mechanism computes an attention weight distribution, $\boldsymbol{\alpha}_t \in \mathbb{R}^{T_h}$, over the audio representations $\mathbf{H}$. This distribution is calculated as:
\[
\boldsymbol{\alpha}_t = \text{softmax} \left( \frac{\mathbf{q}_t \mathbf{K}^\top}{\sqrt{d_k}} \right)
\]
where $\mathbf{q}_t$ is the decoder’s query vector for token $y_t$, and $\mathbf{K}$ are the key vectors derived from $\mathbf{H}$. The weight vector $\boldsymbol{\alpha}_t$ indicates which audio frames the model considers most relevant for transcribing token $y_t$. By feeding our synthesized audio to $\mathcal{M}_{\text{ASR}}$ and using the ground-truth text for teacher-forcing, we extract the full attention map $\mathbf{A} \in \mathbb{R}^{T_y \times T_h}$, where each row $\boldsymbol{\alpha}_t = \mathbf{A}_{t,:}$. This map provides the basis for our quality metrics.

\paragraph{Attention Purity.} A clearly articulated word should correspond to a compact and acoustically focused segment of the audio. A high-quality synthesis should therefore elicit a sharp, unimodal attention distribution from the ASR model, whereas poorly formed speech would result in a diffuse or scattered attention. \par
We define \textit{Attention Purity} as the amount of attention mass concentrated around the attention peak. First, we identify the audio frame with the maximum attention for text token $y_t$: $j_t^* = \arg\max_j A_{t,j}$. The purity reward, $\mathcal{R}_{\text{purity}}$, is the sum of attention weights within a small window of width $W$ centered at $j_t^*$:
\[
\mathcal{R}_{\text{purity}}(y_t) = \sum_{j=j_t^* - W/2}^{j_t^* + W/2} A_{t,j}
\]
Here, $W$ is a hyperparameter representing a short duration. It is noted that this window is not intended to encompass the duration of an entire word, but rather to measure the "sharpness" of the attention distribution around its highest peak.

\paragraph{Alignment Monotonicity.} For fluent speech, the temporal alignment between the spoken words and the audio signal must be strictly monotonic. The focus of attention should consistently progress forward in time. A backward or stalled movement of the attention peak is a strong indicator of a synthesis artifact, such as a stutter, unnatural pause, or prosodic break, which confuses the ASR model’s alignment process. \par
We measure the forward progression of the attention peak location $j_t^*$ relative to the previous token’s peak $j_{t-1}^*$. The \textit{alignment monotonicity} reward, $\mathcal{R}_{\text{mono}}$, is defined as:
\[
\mathcal{R}_{\text{mono}}(y_t) = \tanh \left( \beta (j_t^* - j_{t-1}^*) \right)
\]
where $\beta$ is a scaling hyperparameter. The $\tanh$ function bounds the reward, rewarding forward progression ($j_t^* > j_{t-1}^*$) and penalizing regressions or stalls. For the first token ($t=1$), the score is set to a neutral value of 0.
\begin{algorithm}[t]
\caption{ASR-Guided Group Relative Policy Optimization for TTS}
\label{alg:main}
\begin{algorithmic}[1]
\STATE \textbf{Initialize:} TTS policy $\mathcal{M}_{\theta}$, pre-trained ASR model $\mathcal{M}_{\text{ASR}}$, group size $N$, RL weight $\gamma$.
\FOR{each training iteration}
\STATE Sample a mini-batch of $(\mathbf{y}, \mathbf{a}_{\text{prompt}})$ from the training dataset $\mathcal{D}$.
\STATE \textit{// Supervised pre-training step}
\STATE Compute supervised loss $\mathcal{L}_{\text{train}}$ using the ground-truth acoustic tokens.
\STATE \textit{// RL fine-tuning step}
\FOR{each $(\mathbf{y}, \mathbf{a}_{\text{prompt}})$ in the mini-batch}
\STATE \textit{// 1. Sampling}
\STATE Generate a group of $N$ samples $\{\mathbf{a}^{(1)}, \dots, \mathbf{a}^{(N)}\}$ using the current policy $\pi_{\theta}$.
\STATE \textit{// 2. Reward Calculation}
\FOR{$n=1$ to $N$}
\STATE Synthesize waveform $\mathbf{x}^{(n)} = \mathcal{D}_{\text{codec}}(\mathbf{a}^{(n)})$.
\STATE Compute word-level rewards $\mathcal{R}^{(n)} = (\mathcal{R}(y_1)^{(n)}, \dots)$ using $\mathcal{M}_{\text{ASR}}(\mathbf{x}^{(n)}, \mathbf{y})$.
\ENDFOR
\STATE \textit{// 3. Advantage Calculation}
\STATE Compute word-level advantages $\{A(y_i)^{(n)}\}$ for all samples by de-meaning rewards within the group.
\ENDFOR
\STATE \textit{// 4. Loss Calculation and Policy Update}
\STATE Compute the RL loss $\mathcal{L}_{\text{RL}}$ using the calculated advantages.
\STATE Compute the total loss $\mathcal{L}_{\text{total}} = \mathcal{L}_{\text{train}} + \gamma \mathcal{L}_{\text{RL}}$.
\STATE Update TTS model parameters $\theta$ by descending the gradient: $\theta \leftarrow \theta - \eta \nabla_{\theta} \mathcal{L}_{\text{total}}$.
\ENDFOR
\end{algorithmic}
\end{algorithm}
\paragraph{Combined Word-Level Reward.}

To provide a holistic quality score for each synthesized word, we combine our two metrics into a single word-level reward, $\mathcal{R}(y_t)$. This reward captures both local articulatory clarity (purity) and global prosodic fluency (monotonicity). The final reward is a weighted sum:
\[
\mathcal{R}(y_t) = \lambda_{\text{purity}} \mathcal{R}_{\text{purity}}(y_t) + \lambda_{\text{mono}} \mathcal{R}_{\text{mono}}(y_t)
\]
where $\lambda_{\text{purity}}$ and $\lambda_{\text{mono}}$ are scalar weights that balance the contribution of each component. This reward signal is then used to optimize our TTS model $\mathcal{M}_{\text{TTS}}$ within a reinforcement learning framework.

\subsection{Policy Optimization}
We refine the baseline TTS model $\mathcal{M}_\theta$ by treating it as a policy in a reinforcement learning framework. The optimization process uses the word-level reward signal from the ASR model to directly guide the policy towards generating higher-quality speech. Our approach is inspired by group-based optimization methods, which have proven effective in generation tasks.

\paragraph{Group-Based Sampling and Evaluation.}
The core of the method involves comparing a group of candidate samples generated from the same input. For a given text $\mathbf{y}$ and acoustic prompt $\mathbf{a}_{\text{prompt}}$, we use the current TTS policy, $\pi_\theta(\cdot \mid \mathbf{y}, \mathbf{a}_{\text{prompt}})$, to generate a group of $N$ distinct acoustic sequences, $\{\mathbf{a}^{(1)}, \mathbf{a}^{(2)}, \dots, \mathbf{a}^{(N)}\}$, by employing nucleus or temperature-controlled sampling.

Each sample $\mathbf{a}^{(n)}$ in the group is then evaluated by the ASR model $\mathcal{M}_{\text{ASR}}$ to obtain a sequence of word-level rewards, $\mathcal{R}^{(n)} = (\mathcal{R}(y_1^{(n)}), \dots, \mathcal{R}(y_{T_y}^{(n)}))$, as defined in the previous section.

\paragraph{Word-Level Advantage Function.}
Instead of using the absolute reward values, which can have high variance, we define a fine-grained advantage function that normalizes rewards within the generated group. The advantage of the pronunciation of a word $y_i$ in sample $n$ is calculated as its reward relative to the average reward for that same word across all $N$ samples. This effectively uses the group's average performance as a dynamic baseline.

\[
A(y_i^{(n)}) = \mathcal{R}(y_i^{(n)}) - \frac{1}{N} \sum_{k=1}^{N} \mathcal{R}(y_i^{(k)})
\]

A positive advantage $A(y_i^{(n)})$ indicates that the word's rendering in sample $n$ is superior to the group's average, while a negative value indicates it is inferior. This fine-grained, word-level advantage signal is crucial for targeted policy updates.

\paragraph{Optimization Objective.}
Our goal is to update the policy $\pi_\theta$ to increase the likelihood of acoustic sequences that correspond to positive advantages. We formulate a policy gradient-style objective function, where the word-level advantage modulates the learning signal for each corresponding acoustic token. Let $w(t)$ be a function that maps an acoustic token index $t$ to its corresponding word index $i$. The RL loss is defined as:

\[
\mathcal{L}_{\text{RL}} = - \mathbb{E} \left[ \sum_{n=1}^{N} \sum_{t=1}^{T_a^{(n)}} A(y_{w(t)}^{(n)}) \log \pi_\theta(a_t^{(n)} \mid \mathbf{a}_{<t}^{(n)}, \mathbf{y}, \mathbf{a}_{\text{prompt}}) \right]
\]

\textbf{Intuition:} This objective function directly steers the generative process. When a word is synthesized well (positive advantage), the loss encourages the model to increase the probability of the acoustic tokens that produced it. Conversely, if a word is synthesized poorly (negative advantage), the model is discouraged from generating those specific acoustic tokens in the future. This provides a tight feedback loop that addresses specific articulatory and prosodic failures, pushing the overall distribution of the policy towards generating speech that is consistently rated higher by the ASR-based evaluator. To stabilize training, we add a small weight of KL constrain $\mathcal{L}_{\text{kl}}$ that is computed by the logits of a frozen reference model $\pi_{\text{ref}}$:

\[
\mathcal{L}_{\text{total}} =  \mathcal{L}_{\text{RL}} + \gamma \mathcal{L}_{\text{kl}}(\pi_\text{ref} || \pi_{\theta}) 
\]

\begin{table*}[!t]
\centering
\caption{Main experimental results comparing our proposed W3AR method with the baseline CoSyVoice model. We report both objective and subjective metrics on In-Domain and Out-of-Domain test sets. The arrows (↑/↓) indicate whether higher or lower values are better. For MOS scores, we report the 95\% confidence intervals. Our method demonstrates significant improvements across all metrics, especially on the out-of-domain data.}
\label{tab:main_results}
\resizebox{\textwidth}{!}{
\begin{tabular}{llccccc}
\toprule
\multirow{2}{*}{\textbf{Dataset}} & \multirow{2}{*}{\textbf{Method}} & \multicolumn{3}{c}{\textbf{Objective Metrics}} & \multicolumn{2}{c}{\textbf{Subjective Metrics}} \\
\cmidrule(lr){3-5} \cmidrule(lr){6-7}
& & \textbf{WER} ($\downarrow$) & \textbf{SECS} ($\uparrow$) & \textbf{BC} (\%) ($\downarrow$) & \textbf{MOS-N} ($\uparrow$) & \textbf{MOS-S} ($\uparrow$) \\
\midrule
\multirow{3}{*}{In-Domain (LibriTTS)} & Baseline & 5.25 & 0.69 &10.9 & 4.07 $\pm$ 0.06 & 4.21 $\pm$ 0.05 \\
& W3AR & \textbf{3.21} & \textbf{0.71} &  \textbf{4.71} &\textbf{4.38} $\pm$ 0.05 & \textbf{4.32} $\pm$ 0.04 \\
& GroundTruth & 2.19 & - &  - &  - \\
\midrule
\multirow{3}{*}{Out-of-Domain (Emilia/GigaSpeech)} & Baseline & 8.92 & 0.66 &15.4 & 3.81 $\pm$ 0.07 & 4.12 $\pm$ 0.06 \\
&  W3AR & \textbf{4.54} & \textbf{0.69}&\textbf{8.14} & \textbf{4.15} $\pm$ 0.06 & \textbf{4.21} $\pm$ 0.05 \\ 
& GroundTruth & 3.87 & - &  - &  - \\
\bottomrule
\end{tabular}
}
\end{table*}
\section{Experiment Setup}
\subsection{Models and Dataset}
Our primary experiments optimized \textbf{CosyVoice}~\cite{du2024cosyvoice}, a robust TTS model trained on an extensive dataset of approximately 170k hours of audio. To rigorously validate the generality of our proposed method, we also extend our optimization to include other representative TTS architectures, namely \textbf{VoiceCraft}~\cite{peng2024voicecraft} and \textbf{MaskGCT}~\cite{wang2024maskgct}. It is noted that the architectural distinctions among these models: CosyVoice and MaskGCT both employ an autoregressive prediction of single-layer semantic tokens, subsequently reconstructing speech via flow-matching and non-autoregressive methods, respectively. In contrast, VoiceCraft directly predicts multi-layer audio codecs to achieve speech synthesis. For the ASR model, we choose Whisper-large-v2 due to its popularity. \par
Given that our optimization method operates without the need for supervised data, we utilized text from LibriTTS~\cite{zen2019libritts}. For prompts, we sampled 2,500 short utterances, ranging from 3 to 6 seconds in length, from the LibriTTS training set: 2,000 for optimization and 500 for evaluation. Furthermore, because CoSyVoice's training data includes LibriTTS, we extracted an additional 2,500 speech samples from Emilia~\cite{he2024emilia} and GigaSpeech~\cite{chen2021gigaspeech} to serve as an out-of-domain speaker library. \par
\subsection{Training and Evaluation}
\paragraph{Training Details.} Our autoregressive TTS model, which was pre-trained on a large-scale multi-speaker corpus, is fine-tuned using our proposed ASR-guided policy optimization method. The optimization is performed using the AdamW optimizer with a learning rate of $2 \times 10^{-5}$, $\beta_1 = 0.9$, $\beta_2 = 0.98$, and an epsilon of $10^{-9}$. We employ a cosine learning rate decay schedule with a warm-up phase of 2,000 steps. The model is trained using 2 NVIDIA A100 GPUs. For the core of our proposed algorithm, we set the group size for candidate generation to $N = 8$. The total loss is a combination of the KL loss and the RL loss, weighted by $\gamma = 0.1$. The word-level reward function is calculated with its own set of hyperparameters: the attention purity window is set to $W = 6$, the alignment monotonicity scaling factor is set to $\beta = 0.1$, and the two reward components are balanced with equal weights, where $\lambda_{\text{purity}} = 0.5$ and $\lambda_{\text{mono}} = 0.5$. \par
\paragraph{Objective Metrics.} 
To evaluate our system objectively, we assess speaker similarity using the speaker embedding cosine similarity (SECS) metric, computed via pre-trained speaker verification models~\footnote{https://github.com/microsoft/UniSpeech}. Robustness is measured using word error rate (WER), with transcripts generated by Whisper-medium to compare with other works. In terms of speech naturalness, we utilize a neural-network-based estimator to predict mean opinion scores (UTMOS)~\footnote{https://github.com/sarulab-speech/UTMOS22}. Following prior work~\cite{chen2024enhancing}, we also report the bad case ratio (BC), defined as the percentage of samples with either UTMOS below 3 or WER above 20\%, to reflect model robustness across varying conditions. \par
\paragraph{Subjective Metrics.} 
We use the naturalness mean opinion score (MOS-N) and similarity mean opinion score (MOS-S) to evaluate the naturalness of the generated samples from 20 English native speakers. To evaluate the perceived quality of the synthesized audio, we conducted two subjective listening tests. First, the naturalness and similarity of 100 samples are assessed using a 5-point Mean Opinion Score (MOS) scale, where a rating of 1 indicated ``very unnatural" and 5 indicated ``completely natural", a rating of 1 indicated ``very similar" and 5 indicated ``completely same". Additionally, a paired-comparison AB test was performed to determine relative preference. In this test, participants listened to 100 pairs of samples generated from identical input text by two competing models and were instructed to select the sample they perceived as more natural. A ``no preference" (tie) option was provided for cases where the samples were not clearly distinguishable.

\paragraph{Baselines.} We reproduce the following baselines to optimize CosyVoice:
\begin{itemize}
    \item Speech-Align~\cite{zhang2024speechalign}. An utterance-level TTS optimization method that employs ground-truth speech as positive samples, while generated acoustic sequence is negative. This method needs the speech label for optimization, leading to unfair comparison.
    \item UNO~\cite{chen2024enhancing}: An utterance-level TTS optimization method that also consider the uncertainty in MOS estimation. Since the related model is not open-sourced, we skip the uncertainty coefficient with constant. 
    \item FPO~\cite{yao2025fine}: The first work that consider fine-grained reward in TTS optimization. It analyzes the types of issues in generated samples, categorize them into two groups, and propose a selective training loss strategy to optimize preferences based on issue type.
\end{itemize}

\section{Result and Analysis}
To comprehensively evaluate the effectiveness, robustness, and generality of our proposed method, W3AR, we conduct a series of extensive experiments. Our evaluation is structured into three main parts. First, we present the main results by applying W3AR to the state-of-the-art CoSyVoice model, comparing it against the baseline on both in-domain and challenging out-of-domain speakers to validate its core performance and generalization capabilities. Second, we conduct detailed ablation studies to dissect the contribution of each key component within our framework, and compere W3AR with other competetive TTS optimization methods. Finally, we perform generalization and visualization analyses to demonstrate the model-agnostic nature and provide intuitive insights into our approach. 

\begin{figure}[]
\centering
\includegraphics[width=0.9\columnwidth]{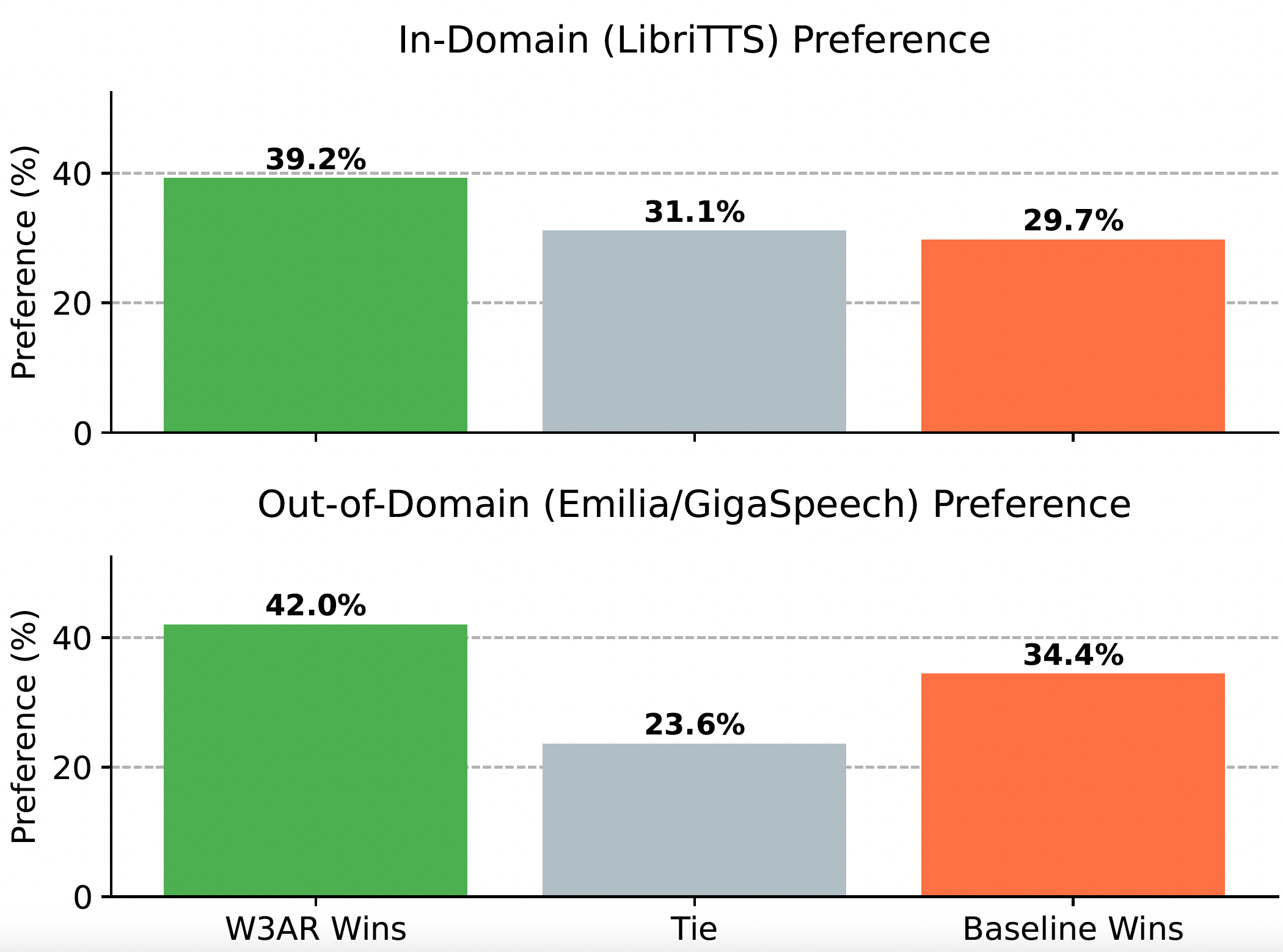} 
\caption{AB test results of W3AR (Ours) and Baseline (CosyVoice) from human listeners}
\label{fig2}
\end{figure}

\subsection{Main Result}
The qualitative results are presented in Table~\ref{tab:main_results}, demonstrating the clear superiority and robustness of our proposed W3AR method. On the In-Domain (LibriTTS) test set, W3AR significantly reduces the Word Error Rate (WER) by 38.9\% relative to the baseline (from 5.25 to 3.21), indicating a substantial improvement in articulatory precision. This is corroborated by subjective evaluations, where our method achieves a notable increase in both Mean Opinion Scores for Naturalness (MOS-N) and Speaker Similarity (MOS-S). The effectiveness of W3AR is particularly pronounced on the more challenging Out-of-Domain (Emilia/GigaSpeech) set. Here, our method nearly halves the WER (from 8.92 to 4.54) and brings the MOS-N for these unseen speakers (4.15) to a level on par with that of the baseline on in-domain data. This highlights our method's strong generalization capability and its effectiveness in mitigating quality degradation for novel voices. \par
To further validate these findings from a human-centric perspective, we conducted a head-to-head AB preference test, with the results displayed in Figure 2. For In-Domain speakers, listeners showed a clear preference for speech generated by W3AR, selecting it in 39.2\% of comparisons, compared to only 29.7\% for the baseline. This preference becomes even more decisive for Out-of-Domain speakers, where W3AR's win rate increases to 42.0\%. Critically, the percentage of "Tie" results drops from 31.1\% in the in-domain scenario to 23.6\% in the out-of-domain one. This reduction suggests that the quality improvements offered by W3AR are not only significant but also more readily and consistently perceivable by human listeners on more challenging voices, where the baseline model is more likely to produce discernible artifacts.

\begin{table*}[!t]
\centering
\caption{Ablation study of our proposed W3AR method and comparison with recent TTS optimization baselines. We report objective metrics: WER, SECS, and UTMOS. The results demonstrate that each component of W3AR contributes positively and that W3AR is highly competitive, especially in out-of-domain generalization.}
\vspace{-0.2cm}
\label{tab:ablation_and_baselines}
\begin{tabular}{lcccccc}
\toprule
\multirow{2}{*}{\textbf{Method}} & \multicolumn{3}{c}{\textbf{In-Domain (LibriTTS)}} & \multicolumn{3}{c}{\textbf{Out-of-Domain (Emilia/GigaSpeech)}} \\
\cmidrule(lr){2-4} \cmidrule(lr){5-7}
& WER ($\downarrow$) & SECS ($\uparrow$) & UTMOS ($\uparrow$) & WER ($\downarrow$) & SECS ($\uparrow$) & UTMOS ($\uparrow$) \\
\midrule
CosyVoice (Baseline)
& 5.25 & 0.69 & 3.91 & 8.92 & 0.66 & 3.70 \\
\midrule
\multicolumn{7}{c}{\textit{Ablation Study of W3AR Components}} \\
\midrule
\textbf{W3AR}
& 3.21 & \textbf{0.71} & \textbf{4.10} & \textbf{4.54} & \textbf{0.69} & \textbf{3.99} \\
\quad w/o Purity Reward
& 4.15 & 0.69 & 4.02 & 5.62 & \textbf{0.69} & 3.83 \\
\quad w/o Monotonicity Reward
& 4.98 & 0.71 & 3.95 & 5.98 & 0.68 & 3.75 \\
\quad w/o Group-Relative Opt.
& 4.41 & 0.69 & 3.90 & 7.23 & 0.66 & 3.79 \\
\midrule
\multicolumn{7}{c}{\textit{Comparison with other optimization methods}} \\
\midrule
SpeechAlign~\cite{zhang2024speechalign}
& 3.80 & 0.70 & 4.02 & 5.90 & 0.68 & 3.82 \\
UNO~\cite{chen2024enhancing}
& 3.92 & 0.69 & 3.98 & 6.72 & \textbf{0.69} & 3.70 \\
FPO~\cite{yao2025fine}
& \textbf{3.15 }& 0.68 & 4.05 & 5.94 & 0.66 & 3.78 \\
\bottomrule
\end{tabular}
\vspace{-0.2cm}
\end{table*}

\subsection{Ablation and Comparison}
To validate the contributions of our proposed components, we conducted a thorough ablation study, with results presented in Table~\ref{tab:ablation_and_baselines}. The full W3AR model demonstrates a significant performance gain over the baseline, establishing a strong reference for comparison. The removal of either the Purity or Monotonicity rewards results in a clear degradation across all metrics, particularly in WER and the predicted UTMOS, which confirms their complementary roles in enhancing articulatory clarity and prosodic fluency. Most notably, disabling the Group-Relative Optimization strategy leads to the most severe performance drop, especially in the challenging out-of-domain scenario where the WER increases from 4.54 to 7.23. This result underscores the critical function of our optimization strategy in ensuring stable and effective policy updates for robust generalization \par
We further benchmark W3AR against several recent TTS optimization methods to contextualize its performance. As shown in Table~\ref{tab:ablation_and_baselines}, our W3AR framework is highly competitive. While FPO achieves a marginally lower WER on the in-domain set, our W3AR model attains superior speaker similarity (SECS) and the highest perceived quality (UTMOS). More importantly, W3AR demonstrates a significant advantage in generalization. It achieves a substantially lower WER (4.54) and a higher UTMOS (3.95) on the out-of-domain set compared to all other methods, including FPO (5.94 WER, 3.78 UTMOS). This superior performance on unseen speakers positions W3AR as a more robust and practical solution for improving the reliability of zero-shot TTS systems.

\subsection{Cross-model Generalization Result}
\begin{table}[t]
\centering
\caption{Cross-model generalization results. We apply our W3AR optimization method to two other representative TTS models, VoiceCraft and MaskGCT. Objective metrics are reported on the LibriTTS test set. The results show that W3AR consistently improves performance across different model architectures, demonstrating its universality.}
\vspace{-0.1cm}
\label{tab:generalization}
\resizebox{\columnwidth}{!}{%
\begin{tabular}{llcc}
\toprule
\textbf{Base Model} & \textbf{Method} & \textbf{WER} ($\downarrow$) & \textbf{UTMOS} ($\uparrow$) \\
\midrule
\multirow{2}{*}{VoiceCraft} & Baseline & 8.55 & 3.62 \\
& + W3AR (Ours) & \textbf{4.98} & \textbf{3.95} \\
\midrule
\multirow{2}{*}{MaskGCT} & Baseline & 2.92 & 4.11 \\
& + W3AR (Ours) & \textbf{2.71} & \textbf{4.23} \\
\bottomrule
\end{tabular}%
}
\vspace{-0.1cm}
\end{table}

Model-Agnostic Generalization. To verify the universality of our proposed W3AR framework, we applied it to two additional state-of-the-art TTS models with distinct architectures: VoiceCraft and MaskGCT. The results, presented in Table 3, confirm that our method is not limited to a single model family. When applied to VoiceCraft, W3AR yields a dramatic relative WER reduction of 41.7\% (from 8.55 to 4.98) and a corresponding significant increase in the predicted quality score (UTMOS). Furthermore, despite the already strong performance of the MaskGCT baseline (2.92 WER), our method still manages to reduce its error rate and improve its UTMOS score. These experiments demonstrate that W3AR functions as a versatile and effective post-optimization layer, capable of enhancing a wide range of modern TTS systems regardless of their underlying generative mechanism.

\section{Conclusion}
In this paper, we address the challenge of fine-grained optimization for autoregressive TTS by introducing W3AR, a novel framework that uses a pre-trained ASR model to provide word-level rewards. W3AR derives these rewards by analyzing the Attention Purity and Alignment Monotonicity of an ASR cross-attention map, effectively assessing both articulatory clarity and prosodic fluency. This signal then guides a stable group-relative policy optimization process to directly correct specific generation artifacts. Our experiments show W3AR significantly improves state-of-the-art TTS models, demonstrating superior performance and generalization on both out-of-domain speakers and across different model architectures. Our work establishes a powerful paradigm for generative AI refinement, demonstrating that the internal "perception" of one expert model can be distilled into a fine-grained, interpretable reward to systematically enhance the quality and robustness of another.

\newpage
\bibliography{aaai2026}

@article{cao2025survey,
  title={A survey of ai-generated content (aigc)},
  author={Cao, Yihan and Li, Siyu and Liu, Yixin and Yan, Zhiling and Dai, Yutong and Yu, Philip and Sun, Lichao},
  journal={ACM Computing Surveys},
  volume={57},
  number={5},
  pages={1--38},
  year={2025},
  publisher={ACM New York, NY}
}

@inproceedings{qian2014training,
  title={On the training aspects of deep neural network (DNN) for parametric TTS synthesis},
  author={Qian, Yao and Fan, Yuchen and Hu, Wenping and Soong, Frank K},
  booktitle={2014 IEEE International Conference on Acoustics, Speech and Signal Processing (ICASSP)},
  pages={3829--3833},
  year={2014},
  organization={IEEE}
}

@article{wang2023neural,
  title={Neural codec language models are zero-shot text to speech synthesizers},
  author={Wang, Chengyi and Chen, Sanyuan and Wu, Yu and Zhang, Ziqiang and Zhou, Long and Liu, Shujie and Chen, Zhuo and Liu, Yanqing and Wang, Huaming and Li, Jinyu and others},
  journal={arXiv preprint arXiv:2301.02111},
  year={2023}
}

@article{chen2024vall,
  title={Vall-e 2: Neural codec language models are human parity zero-shot text to speech synthesizers},
  author={Chen, Sanyuan and Liu, Shujie and Zhou, Long and Liu, Yanqing and Tan, Xu and Li, Jinyu and Zhao, Sheng and Qian, Yao and Wei, Furu},
  journal={arXiv preprint arXiv:2406.05370},
  year={2024}
}

@article{wang2024maskgct,
  title={Maskgct: Zero-shot text-to-speech with masked generative codec transformer},
  author={Wang, Yuancheng and Zhan, Haoyue and Liu, Liwei and Zeng, Ruihong and Guo, Haotian and Zheng, Jiachen and Zhang, Qiang and Zhang, Xueyao and Zhang, Shunsi and Wu, Zhizheng},
  journal={arXiv preprint arXiv:2409.00750},
  year={2024}
}

@inproceedings{popov2021grad,
  title={Grad-tts: A diffusion probabilistic model for text-to-speech},
  author={Popov, Vadim and Vovk, Ivan and Gogoryan, Vladimir and Sadekova, Tasnima and Kudinov, Mikhail},
  booktitle={International conference on machine learning},
  pages={8599--8608},
  year={2021},
  organization={PMLR}
}

@article{du2024cosyvoice,
  title={Cosyvoice: A scalable multilingual zero-shot text-to-speech synthesizer based on supervised semantic tokens},
  author={Du, Zhihao and Chen, Qian and Zhang, Shiliang and Hu, Kai and Lu, Heng and Yang, Yexin and Hu, Hangrui and Zheng, Siqi and Gu, Yue and Ma, Ziyang and others},
  journal={arXiv preprint arXiv:2407.05407},
  year={2024}
}

@article{du2024cosyvoice2,
  title={Cosyvoice 2: Scalable streaming speech synthesis with large language models},
  author={Du, Zhihao and Wang, Yuxuan and Chen, Qian and Shi, Xian and Lv, Xiang and Zhao, Tianyu and Gao, Zhifu and Yang, Yexin and Gao, Changfeng and Wang, Hui and others},
  journal={arXiv preprint arXiv:2412.10117},
  year={2024}
}

@article{xin2024rall,
  title={Rall-e: Robust codec language modeling with chain-of-thought prompting for text-to-speech synthesis},
  author={Xin, Detai and Tan, Xu and Shen, Kai and Ju, Zeqian and Yang, Dongchao and Wang, Yuancheng and Takamichi, Shinnosuke and Saruwatari, Hiroshi and Liu, Shujie and Li, Jinyu and others},
  journal={arXiv preprint arXiv:2404.03204},
  year={2024}
}

@article{neekhara2024improving,
  title={Improving robustness of llm-based speech synthesis by learning monotonic alignment},
  author={Neekhara, Paarth and Hussain, Shehzeen and Ghosh, Subhankar and Li, Jason and Valle, Rafael and Badlani, Rohan and Ginsburg, Boris},
  journal={arXiv preprint arXiv:2406.17957},
  year={2024}
}

@article{zhang2024speechalign,
  title={Speechalign: Aligning speech generation to human preferences},
  author={Zhang, Dong and Li, Zhaowei and Li, Shimin and Zhang, Xin and Wang, Pengyu and Zhou, Yaqian and Qiu, Xipeng},
  journal={Advances in Neural Information Processing Systems},
  volume={37},
  pages={50343--50360},
  year={2024}
}

@article{yao2025fine,
  title={Fine-grained Preference Optimization Improves Zero-shot Text-to-Speech},
  author={Yao, Jixun and Yang, Yuguang and Pan, Yu and Feng, Yuan and Ning, Ziqian and Ye, Jianhao and Zhou, Hongbin and Xie, Lei},
  journal={arXiv preprint arXiv:2502.02950},
  year={2025}
}

@article{ben2024attend,
  title={Attend first, consolidate later: On the importance of attention in different llm layers},
  author={Ben-Artzy, Amit and Schwartz, Roy},
  journal={arXiv preprint arXiv:2409.03621},
  year={2024}
}

@article{hu2024self,
  title={Self-taught recognizer: Toward unsupervised adaptation for speech foundation models},
  author={Hu, Yuchen and Chen, Chen and Yang, Chao-Han and Qin, Chengwei and Chen, Pin-Yu and Chng, Eng-Siong and Zhang, Chao},
  journal={Advances in Neural Information Processing Systems},
  volume={37},
  pages={29566--29594},
  year={2024}
}

@article{peng2024voicecraft,
  title={Voicecraft: Zero-shot speech editing and text-to-speech in the wild},
  author={Peng, Puyuan and Huang, Po-Yao and Li, Shang-Wen and Mohamed, Abdelrahman and Harwath, David},
  journal={arXiv preprint arXiv:2403.16973},
  year={2024}
}

@article{zeghidour2021soundstream,
  title={Soundstream: An end-to-end neural audio codec},
  author={Zeghidour, Neil and Luebs, Alejandro and Omran, Ahmed and Skoglund, Jan and Tagliasacchi, Marco},
  journal={IEEE/ACM Transactions on Audio, Speech, and Language Processing},
  volume={30},
  pages={495--507},
  year={2021},
  publisher={IEEE}
}

@inproceedings{wu2023audiodec,
  title={Audiodec: An open-source streaming high-fidelity neural audio codec},
  author={Wu, Yi-Chiao and Gebru, Israel D and Markovi{\'c}, Dejan and Richard, Alexander},
  booktitle={ICASSP 2023-2023 IEEE International Conference on Acoustics, Speech and Signal Processing (ICASSP)},
  pages={1--5},
  year={2023},
  organization={IEEE}
}

@inproceedings{du2025vall,
  title={Vall-t: Decoder-only generative transducer for robust and decoding-controllable text-to-speech},
  author={Du, Chenpeng and Guo, Yiwei and Wang, Hankun and Yang, Yifan and Niu, Zhikang and Wang, Shuai and Zhang, Hui and Chen, Xie and Yu, Kai},
  booktitle={ICASSP 2025-2025 IEEE International Conference on Acoustics, Speech and Signal Processing (ICASSP)},
  pages={1--5},
  year={2025},
  organization={IEEE}
}

@article{wang2025spark,
  title={Spark-tts: An efficient llm-based text-to-speech model with single-stream decoupled speech tokens},
  author={Wang, Xinsheng and Jiang, Mingqi and Ma, Ziyang and Zhang, Ziyu and Liu, Songxiang and Li, Linqin and Liang, Zheng and Zheng, Qixi and Wang, Rui and Feng, Xiaoqin and others},
  journal={arXiv preprint arXiv:2503.01710},
  year={2025}
}

@article{anastassiou2024seed,
  title={Seed-tts: A family of high-quality versatile speech generation models},
  author={Anastassiou, Philip and Chen, Jiawei and Chen, Jitong and Chen, Yuanzhe and Chen, Zhuo and Chen, Ziyi and Cong, Jian and Deng, Lelai and Ding, Chuang and Gao, Lu and others},
  journal={arXiv preprint arXiv:2406.02430},
  year={2024}
}

@article{zheng2023secrets,
  title={Secrets of rlhf in large language models part i: Ppo},
  author={Zheng, Rui and Dou, Shihan and Gao, Songyang and Hua, Yuan and Shen, Wei and Wang, Binghai and Liu, Yan and Jin, Senjie and Liu, Qin and Zhou, Yuhao and others},
  journal={arXiv preprint arXiv:2307.04964},
  year={2023}
}

@article{kaufmann2024survey,
  title={A survey of reinforcement learning from human feedback},
  author={Kaufmann, Timo and Weng, Paul and Bengs, Viktor and H{\"u}llermeier, Eyke},
  year={2024}
}

@article{lee2023rlaif,
  title={Rlaif: Scaling reinforcement learning from human feedback with ai feedback},
  author={Lee, Harrison and Phatale, Samrat and Mansoor, Hassan and Lu, Kellie Ren and Mesnard, Thomas and Ferret, Johan and Bishop, Colton and Hall, Ethan and Carbune, Victor and Rastogi, Abhinav},
  year={2023}
}

@article{wu2023fine,
  title={Fine-grained human feedback gives better rewards for language model training},
  author={Wu, Zeqiu and Hu, Yushi and Shi, Weijia and Dziri, Nouha and Suhr, Alane and Ammanabrolu, Prithviraj and Smith, Noah A and Ostendorf, Mari and Hajishirzi, Hannaneh},
  journal={Advances in Neural Information Processing Systems},
  volume={36},
  pages={59008--59033},
  year={2023}
}

@article{chen2024enhancing,
  title={Enhancing zero-shot text-to-speech synthesis with human feedback},
  author={Chen, Chen and Hu, Yuchen and Wu, Wen and Wang, Helin and Chng, Eng Siong and Zhang, Chao},
  journal={arXiv preprint arXiv:2406.00654},
  year={2024}
}

@inproceedings{tian2025preference,
  title={Preference alignment improves language model-based tts},
  author={Tian, Jinchuan and Zhang, Chunlei and Shi, Jiatong and Zhang, Hao and Yu, Jianwei and Watanabe, Shinji and Yu, Dong},
  booktitle={ICASSP 2025-2025 IEEE International Conference on Acoustics, Speech and Signal Processing (ICASSP)},
  pages={1--5},
  year={2025},
  organization={IEEE}
}

@article{hu2024robust,
  title={Robust zero-shot text-to-speech synthesis with reverse inference optimization},
  author={Hu, Yuchen and Chen, Chen and Wang, Siyin and Chng, Eng Siong and Zhang, Chao},
  journal={arXiv preprint arXiv:2407.02243},
  year={2024}
}

@article{zen2019libritts,
  title={Libritts: A corpus derived from librispeech for text-to-speech},
  author={Zen, Heiga and Dang, Viet and Clark, Rob and Zhang, Yu and Weiss, Ron J and Jia, Ye and Chen, Zhifeng and Wu, Yonghui},
  journal={arXiv preprint arXiv:1904.02882},
  year={2019}
}

@inproceedings{he2024emilia,
  title={Emilia: An extensive, multilingual, and diverse speech dataset for large-scale speech generation},
  author={He, Haorui and Shang, Zengqiang and Wang, Chaoren and Li, Xuyuan and Gu, Yicheng and Hua, Hua and Liu, Liwei and Yang, Chen and Li, Jiaqi and Shi, Peiyang and others},
  booktitle={2024 IEEE Spoken Language Technology Workshop (SLT)},
  pages={885--890},
  year={2024},
  organization={IEEE}
}

@article{chen2021gigaspeech,
  title={Gigaspeech: An evolving, multi-domain asr corpus with 10,000 hours of transcribed audio},
  author={Chen, Guoguo and Chai, Shuzhou and Wang, Guanbo and Du, Jiayu and Zhang, Wei-Qiang and Weng, Chao and Su, Dan and Povey, Daniel and Trmal, Jan and Zhang, Junbo and others},
  journal={arXiv preprint arXiv:2106.06909},
  year={2021}
}

@inproceedings{radford2023robust,
  title={Robust speech recognition via large-scale weak supervision},
  author={Radford, Alec and Kim, Jong Wook and Xu, Tao and Brockman, Greg and McLeavey, Christine and Sutskever, Ilya},
  booktitle={International conference on machine learning},
  pages={28492--28518},
  year={2023},
  organization={PMLR}
}

\end{document}